\documentclass[transmag,letterpaper]{IEEEtran}
\newif\ifpeerreview

\peerreviewfalse

\newcommand{\paperID}{35}

\usepackage{xcolor}
\usepackage[switch]{lineno}
\usepackage{cite}
\usepackage{url}
\usepackage{lipsum}
\usepackage{graphicx}
\usepackage{subfigure}
\usepackage{amsfonts}
\usepackage{amsmath}
\usepackage[ruled,noline,linesnumbered,noend]{algorithm2e}
\usepackage{booktabs}
\usepackage[bookmarks=false]{hyperref}
\usepackage{nopageno}

\DeclareMathOperator*{\argmin}{arg\,min}

\ifCLASSINFOpdf
\else
\fi

\begin{document}

\ifpeerreview
  \linenumbers
  \linenumbersep 5pt\relax
\fi

%
\title{4D X-Ray CT Reconstruction using Multi-Slice Fusion}


\ifpeerreview
\author{Anonymous ICCP 2019 submission \\
Paper ID \paperID}
\else
\author{\IEEEauthorblockN{Soumendu Majee\IEEEauthorrefmark{1},
Thilo Balke\IEEEauthorrefmark{1},
Craig A. J. Kemp\IEEEauthorrefmark{2}, 
Gregery T. Buzzard\IEEEauthorrefmark{3}, and
Charles A. Bouman\IEEEauthorrefmark{1}}
\IEEEauthorblockA{\IEEEauthorrefmark{1}School of Electrical and Computer Engineering,
Purdue University, West Lafayette, IN, USA}
\IEEEauthorblockA{\IEEEauthorrefmark{2}Eli Lilly and Company, Indianapolis, IN, USA}
\IEEEauthorblockA{\IEEEauthorrefmark{3}Department of Mathematics, Purdue University, West Lafayette, IN, USA}

}
\fi

\ifpeerreview
\markboth{Anonymous ICCP 2019 submission ID \paperID}%
{}
\else
\fi

\IEEEtitleabstractindextext{%
\begin{abstract}

There is an increasing need to reconstruct objects in four or more dimensions corresponding to space, time and other independent parameters. 
The best 4D reconstruction algorithms use regularized iterative reconstruction approaches such as model based iterative reconstruction (MBIR), which depends critically on the quality of the prior modeling. 
Recently, Plug-and-Play methods have been shown to be an effective way to incorporate advanced prior models using state-of-the-art denoising algorithms designed to remove additive white Gaussian noise (AWGN). 
However, state-of-the-art denoising algorithms such as BM4D and deep convolutional neural networks (CNNs) are primarily available for 2D and sometimes 3D images. 
In particular, CNNs are difficult and computationally expensive to implement in four or more dimensions, and training may be impossible if there is no associated high-dimensional training data.

In this paper, we present {\em Multi-Slice Fusion}, a novel algorithm for 4D and higher-dimensional reconstruction, based on the fusion of multiple low-dimensional denoisers. 
Our approach uses multi-agent consensus equilibrium (MACE), an extension of Plug-and-Play, as a framework for integrating the multiple lower-dimensional prior models.
We apply our method to the problem of 4D cone-beam X-ray CT reconstruction for Non Destructive Evaluation (NDE) of moving parts.
This is done by solving the MACE equations using lower-dimensional CNN denoisers implemented in parallel on a heterogeneous cluster.
Results on experimental CT data demonstrate that Multi-Slice Fusion can substantially improve the quality of reconstructions relative to traditional 4D priors,
while also being practical to implement and train. 

\end{abstract}

\ifpeerreview
\else
\fi
}

\maketitle
\thispagestyle{empty}
\IEEEdisplaynontitleabstractindextext


\section{Introduction}

Improvements in imaging sensors and computing power have made it possible to solve increasingly difficult reconstruction problems. 
In particular, the dimensionality of reconstruction problems has increased from the traditional~2D and 3D problems representing space
to more difficult 4D or even 5D problems representing space-time and, for example, heart or respiratory phase \cite{5D_huang2014mr,mohan2015timbir}.

These higher-dimensional reconstruction problems pose surprisingly difficult challenges both computationally and perhaps more importantly, in terms of algorithmic design and training due to the curse of dimensionality \cite{ziabariCNN}.
However, the high dimensionality of the reconstruction also presents important opportunities to improve reconstruction quality by exploiting regularity. 
In particular, it has been shown that the temporal resolution of 4D CT imaging can be increased by an order of magnitude by exploiting the time space regularity of objects being imaged \cite{mohan2015timbir,gibbs2015three,mohan20154d,wang2016fast}.
These approaches use model-based iterative reconstruction (MBIR) \cite{kisner2012model,sauer1993local} to enforce regularity using simple space-time prior models.

Recently, it has been demonstrated that Plug-and-Play (P\&P) priors \cite{sreehari2016plug,venkatakrishnan2013plug,sun2018online,kamilov2017plug} can dramatically improve reconstruction quality by enabling the use of state-of-the-art denoisers as prior models in MBIR.
So P\&P has great potential to improve reconstruction quality in 4D CT imaging problems. 
However, state-of-the-art denoisers such as deep Convolutional Neural Networks (CNNs) and BM4D are primarily available for 2D and sometimes 3D images, 
and it is difficult to extend them to higher dimensions \cite{bm3d,bm4d,ziabariCNN}.
In particular, extending CNNs to 4D requires very computationally, and memory intensive 4D convolution applied to 5D feature tensor structures. 
This problem is further compounded by the lack of GPU accelerated routines for 4D convolution from major Deep-Learning frameworks such as Tensorflow, Keras, PyTorch. Currently only 1D, 2D, and 3D convolutions are supported with GPU acceleration. 
Furthermore, 4D CNNs require 4D ground truth data to train the P\&P denoisers, which might be difficult or impossible to obtain.

\begin{figure}[ht]
\centering     
\includegraphics[width=0.50\textwidth]{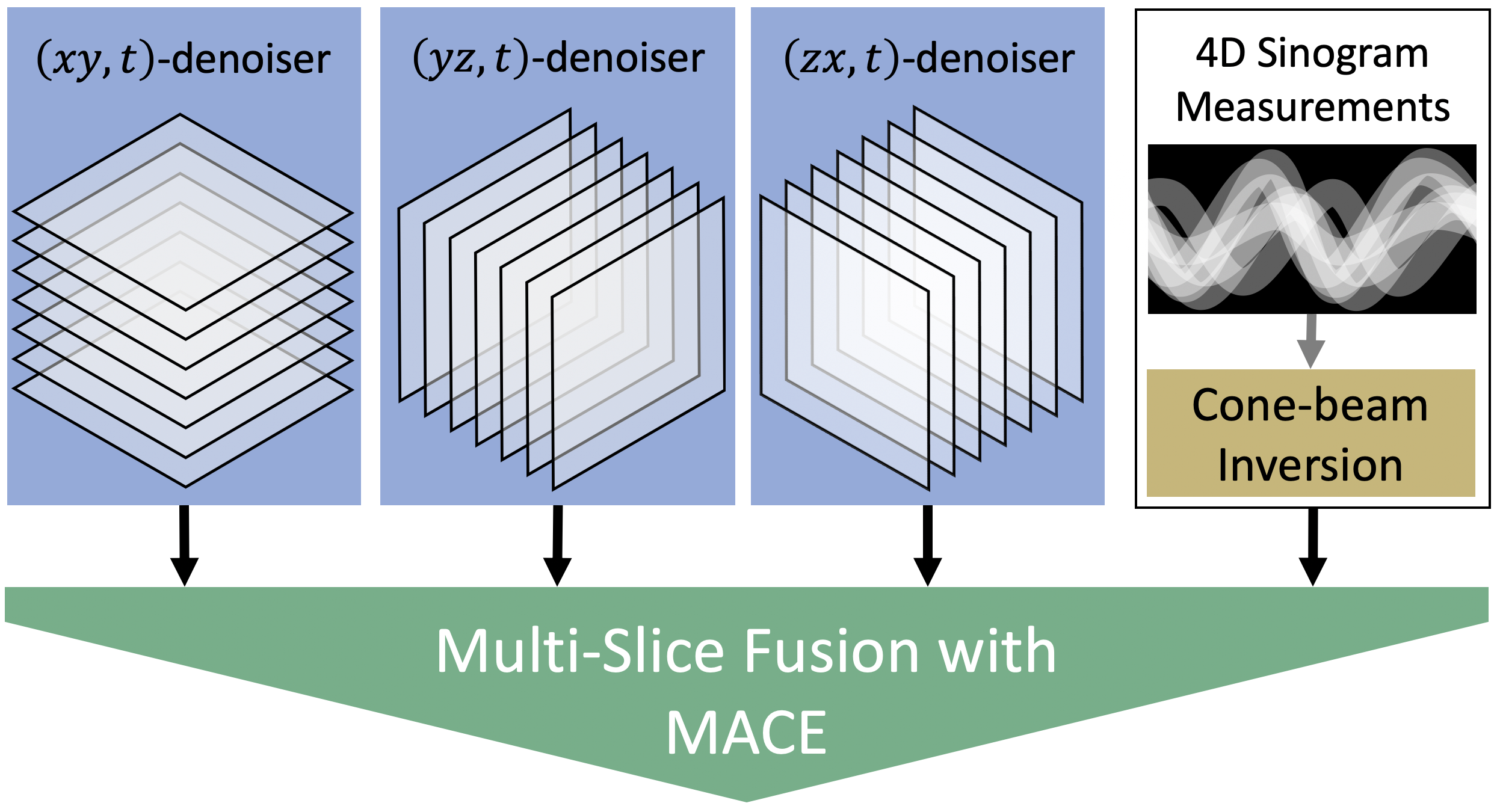}
\caption{Illustration of our Multi-Slice Fusion approach.
We fuse lower dimensional denoisers with the measurement model to produce a 4D regularized reconstruction.}
\label{fig:multislice_fusion}
\end{figure}

In this paper, we present a novel 4D X-ray CT reconstruction algorithm that combines multiple low-dimensional CNN denoisers to implement a highly effective 4D prior model. 
Our approach, which we call {\em Multi-Slice Fusion}, integrates the multiple low-dimensional priors using multi-agent consensus equilibrium (MACE) \cite{buzzard2018plug}.
MACE is an extension of the P\&P framework that formulates the inversion problem using an equilibrium equation---as opposed to an optimization---and allows for the use of multiple prior models and agents.

Figure~\ref{fig:multislice_fusion} illustrates the basic concept of our approach.
Multi-Slice Fusion integrates together three distinct denoisers each of which is designed to remove noise along lower dimensional slices of the 4D object.
When MACE fuses the denoisers it {\em simultaneously} enforces the constraints of each denoising agent,
so the reconstructions are constrained to be smooth in all four dimensions. 
Consequently, Multi-Slice Fusion results in high-quality reconstructions that are practical to train and compute even when the dimensionality of the reconstruction is high.
In our implementation, one MACE agent estimates the cone-beam tomographic inversion.
The remaining 3 agents are CNN denoisers trained to remove additive white Gaussian noise (AWGN) along 2.5D slices of the reconstruction \cite{ziabariCNN,jiang2018denoising,dncnn}.

The MACE solution may be computed using a variety of algorithms, including variants of the Plug-and-Play algorithm based on ADMM or other approaches \cite{sun2018plug,sun2018regularized,venkatakrishnan2013plug,sreehari2016plug}.
We implement our method for the problem of 4D cone-beam X-ray CT reconstruction \cite{balke2018separable} of moving parts, and the MACE solution is computed on a heterogeneous cluster in which different agent updates are computed on different cluster nodes. 
In particular, the algorithm runs in parallel on heterogeneous cluster nodes, with the inversion portion running on CPU nodes and the prior portion running on GPU nodes.

Experimental results on both simulated data and real data obtained from an North Star Imaging X50 X-ray cone-beam scanner indicate that the Multi-Slice Fusion reconstruction using CNN priors can substantially improve reconstructed image quality as compared to MBIR reconstruction using traditional 4D MRF priors.

\section{Problem Formulation}

In X-Ray CT imaging, an object is rotated and several 2D projections (radiographs) of the object are measured for different angles.
The problem is then to reconstruct the 4D array of X-ray attenuation coefficients from these measurements, where 3 dimensions correspond to the spatial dimensions and the fourth dimension corresponds to time.

Let $N_t$ be the number of time-points, and $N_s$ be the number of voxels at each time point of the 4D image.
For each time-point, $n \in \{1,\hdots,N_t\}$, define $y_n \in \mathbb{R}^{M_n} $ to be the vector of sinogram measurements at time $n$, and $x_n \in \mathbb{R}^{N_s}$ to be the 3D volume of X-ray attenuation coefficients for that time-point.
Let us stack all the measurements to form a vector $ Y = [ y_1^\top, .. , y_{N_t}^\top ]^\top \in \mathbb{R}^{M} $ and stack all the 3D volumes to form a vector $ X = [ x_1^\top, \hdots , x_{N_t}^\top  ]^\top \in \mathbb{R}^{N} $, where $N = N_t N_s$ and $M = \sum_{n=1}^{N_t} M_n $.
The 4D reconstruction problem then becomes the task of recovering the 4D image of attenuation coefficients, $X$, from the series of sinogram measurements, $Y$.

The image $X$ can be reconstructed as a maximum a posteriori (MAP) estimate as
\begin{linenomath*}
\begin{equation}\label{eq:map_est}
    X^{*} = \argmin_{X} \left\{ f(X) + R(X) \right\} \ ,
\end{equation}
\end{linenomath*}
where $f(X)$ is the data-fidelity or log-likelihood term and $R(X)$ is the 4D regularizer or prior model for the image, $X$.
The data-fidelity term of the 4D image, $X$, can be written in a separable fashion as
\begin{linenomath*}
\begin{equation}\label{eq:LL_time_expand}
    f(X) = \sum_{n=1}^{N_t} f_n(x_n) \ ,
\end{equation}
\end{linenomath*}
where 
\begin{linenomath*}
\begin{equation}\label{eq:fwdModel}
    f_n(x_n) = \frac{1}{2 \alpha} \| y_n - A_n x_n \|_{\Lambda_n}^2 
\end{equation}
\end{linenomath*}
is the data fidelity term for time $n$, 
where $A_n$ is the system matrix, and $\Lambda_n$ is the weight matrix, and $\alpha$ is a normalizing weight scalar \cite{balke2018separable}.
The values in the diagonal weight matrix, $\Lambda_n$, account for the non-uniform noise variance for each measurement.

If the prior model, $R(X)$, can be expressed analytically like a 4D Markov Random Field (MRF) as in \cite{mohan2015timbir}, then the expression in Equation~(\ref{eq:map_est}) can be minimized iteratively. 
However, in practice, it can be difficult to represent an advanced prior model in the form of a simple cost function $R(X)$ that can be minimized.
Consequently, P\&P methods have been created as a method for representing prior models as denoising operations\cite{sreehari2016plug,venkatakrishnan2013plug}.
More recently, P\&P methods have been generalized to the consesus equilibrium framework as a way to integrate multiple criteria \cite{buzzard2018plug}.

\begin{figure}[ht]
\centering     
\includegraphics[width=0.50\textwidth]{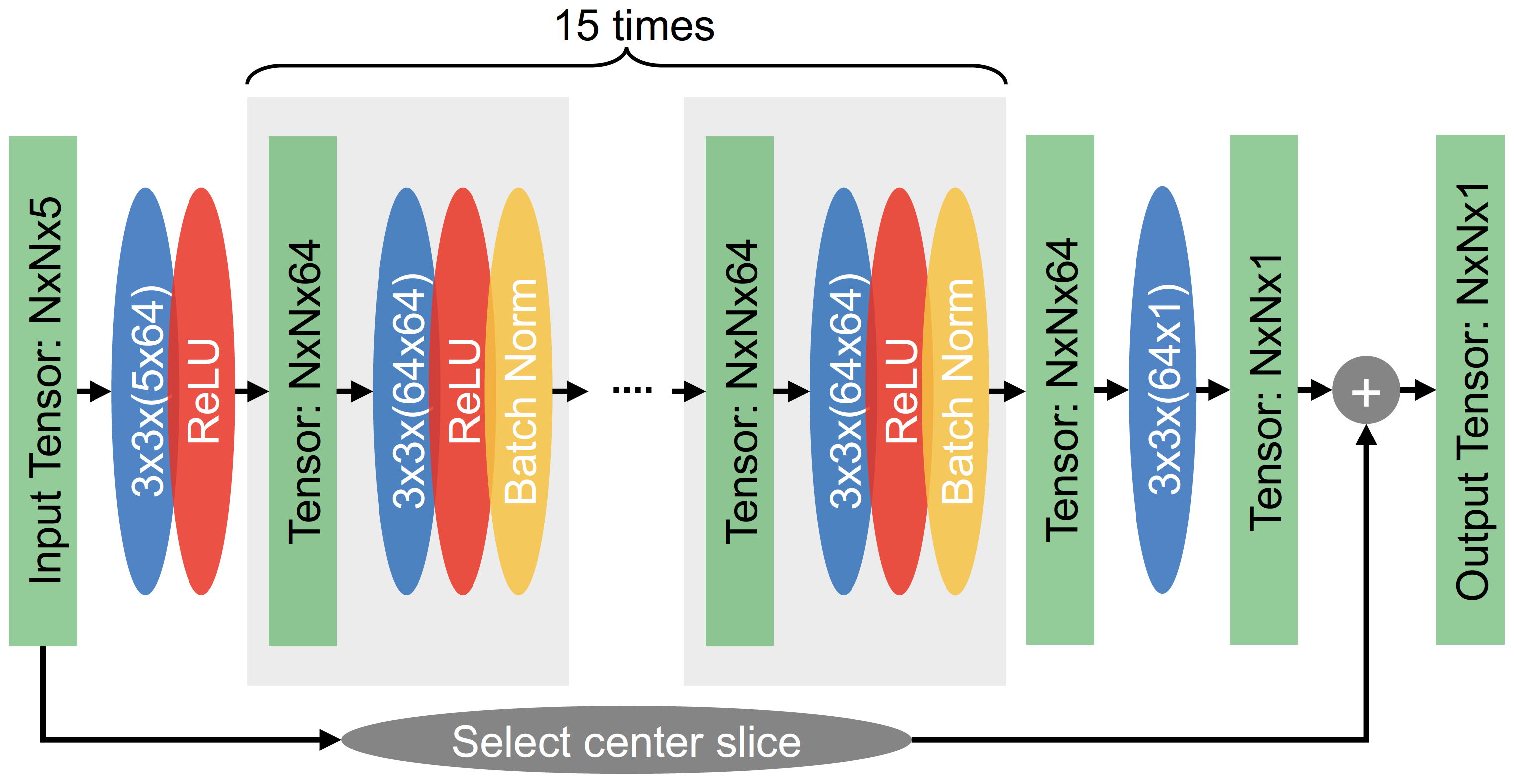}
\caption{Architecture of CNN denoiser. 
Different sizes of input and output necessitate a selection operator for the residual connection.
Each rectangular block denotes a tensor, and each ellipsoidal block denotes an operation.
Blue ellipses specify the shape of the convolution kernel.
}
\label{fig:CNN}
\end{figure}

\section{Fusing Denoisers via Consensus Equilibrium}
\label{sec:MACE_fusion}

We use the MACE framework to fuse multiple denoisers to form a prior model for reconstruction.
This allows us to make a prior model out of denoisers that are not optimization based \cite{buzzard2018plug}.

Let $ H_k $ for $k \in \{1, \hdots, K \}$ be a lower-dimensional denoiser that denoises a full 4D volume by processing the volume ``slice by slice''.
Furthermore, each denoiser, $ H_k $, is trained to remove additive white Gaussian noise (AWGN) of standard deviation $\sigma$ from the image. The denoisers serve as prior models even though a corresponding regularization term, $R(X)$, does not necessarily exist.
The design of these denoisers is described in more detail in section~\ref{sec:denoiser}. Each of the denoisers is a separate agent in the MACE framework.

The data-fidelity is achieved using a proximal map operation, $F$, as another agent. 
The $F$-operator is defined as the proximal map of $f$ as
\begin{linenomath*}
\begin{equation}\label{eq:F_def}
    F(X) = \argmin_{Z \in \mathbb{R}^{N}} \left\{ f(Z) + \frac{\beta}{\sigma^2} \| X-Z \|_2^2 \right\} ,
\end{equation}
\end{linenomath*}
where $\sigma$ is the same as above and the unit-less parameter $\beta$ is used to vary the relative weight of the $F$-operator with with respect to the denoisers and in turn the amount of regularization in the reconstruction can be adjusted.

The operator $F$ enforces fidelity to the measurements while each of the denoisers $H_1$, $ \hdots$, $H_K$ enforces regularity of the image in orthogonal image planes.
Our method imposes a consensus between the operators $F$, $H_1$, $ \hdots$, $H_K$ to achieve a balanced reconstruction that lies at the intersection of the solution space of the measurement model and each of the prior models.

In order to formally state the consensus equilibrium criteria and present the algorithm for computing the solution,
we will need to introduce some additional notation.
First, define the column vector ${ \mathbf{X} = [ X_1^\top, \hdots , X_{K+1}^\top  ]^\top }$ denotes a tall vector $\in \mathbb{R}^{ (K+1) N} $ as the concatination 
of $K+1$ state vectors $X_k\in \mathbb{R}^{N}$ where each state vector represents the input to one of the $K+1$ agents.
Furthermore, let $\mathbf{L}(\mathbf{X} )$ denote the operator formed by the application of all $K+1$ agents on each corresponding state vector.
More precisely, we define the agent operator, $\mathbf{L}$, as
\begin{linenomath*}
\begin{equation}\label{eq:L_operator}
    \mathbf{L}(\mathbf{X}) = \left(\begin{smallmatrix}  
    H_1(X_1) \\ 
    \vdots \\ \\
    H_K(X_{K}) \\ 
    F(X_{K+1})
    \end{smallmatrix}\right) \ .
\end{equation}
\end{linenomath*}
So $\mathbf{L}(\mathbf{X})$ simply denotes the parallel application of each agent to its corresponding state vector.

Next we define the averaging operator, $\mathbf{G}$, as
\begin{linenomath*}
\begin{equation}
    \mathbf{G}(\mathbf{X}) = \left(\begin{smallmatrix} \overline{X} \\
    \overline{X} \\ 
    \vdots \\ \\ 
    \overline{X} \end{smallmatrix}\right),
\end{equation}
\end{linenomath*}
where $\overline{X}$ denotes the weighted average of the state vectors for each agent given by
\begin{linenomath*}
\begin{equation}
    \overline{X} = 
    \frac{1}{2} X_{K+1} + \frac{1}{2 K} \sum_{k=1}^{K} X_k \ .
\end{equation}
\end{linenomath*}
Notice that we choose the weighting so that there is equal weighting of the data-fidelity (i.e., agent $K+1$) and regularization (i.e., agents $1$ to $K$), irrespective of how many denoiser we use to form our prior model.

Using, the above notation, it can be shown \cite{buzzard2018plug,sridhar2018distributed} that when the agents are at consensus, the equilibrium equation
\begin{linenomath*}
\begin{equation}\label{eq:CE}
    \mathbf{L}(\mathbf{X}) = \mathbf{G}(\mathbf{X})
\end{equation}
\end{linenomath*}
holds.
The equilibrium condition of equation~(\ref{eq:CE}) implies that at equilibrium each agent with different inputs produces the same output. 
This output is then the consensus solution that represents a balance point between the forces of each agent. 

From equation~(\ref{eq:CE}) it follows that the consensus solution is the fixed point of the map $\mathbf{T} = (2\mathbf{G}-\mathbf{I})(2\mathbf{L}-\mathbf{I})$ and can found efficiently by using partial update Mann iterations \cite{buzzard2018plug,sridhar2018distributed,sridharDistributed_CT_TCI}.
Partial update Mann iterations use approximate updates for the proximal map operator, $F$.
We define $ \mathbf{\tilde{L}}(\mathbf{X};\mathbf{V}) $ to be the corresponding partial update approximation of  $\mathbf{L}(\mathbf{X})$
\begin{linenomath*}
\begin{equation}\label{eq:L_tilde_operator}
   \mathbf{\tilde{L}}(\mathbf{X};\mathbf{V}) = \left(\begin{smallmatrix} 
    H_1(X_1) \\ 
    \vdots \\ \\ 
    H_K(X_K) \\
    \Tilde{F}(X_{K+1};V_{K+1})
    \end{smallmatrix}\right),
\end{equation}
\end{linenomath*}
where $\mathbf{\tilde{L}}(\mathbf{X};\mathbf{V})$ is the same as $\mathbf{L}(\mathbf{X})$ but uses only an approximation proximal mapping formed by a single update of a iterative optimization method $\Tilde{F}(X_{K+1};V_{K+1})$ initialized by $V_{K+1}$. 
That is, the iterative optimization in the $F$-operator in equation~(\ref{eq:F_def}) is initialized by $V_{K+1}$ and terminated early.

The the partial update Mann iterations for fusing denoisers can be simplified to form Algorithm~\ref{algo:CE_updates}.
The weighing parameter $\rho \in (0,1)$ is used to control speed of convergence.
The initial reconstruction used should be close to the final solution for faster convergence.
Output of a fast conventional algorithm can be used as the initial reconstruction.
We continue the partial update Mann iterations until the differences between state vectors $X_k$ become smaller than a fixed threshold.
We use three passes of Iterative Coordinate Descent (ICD) for computing the approximate proximal mapping, $\Tilde{F}(X_{K+1};V_{K+1})$.

\begin{algorithm}
\label{algo:CE_updates}
\DontPrintSemicolon
\KwIn{Initial Reconstruction: $ X^{(0)} \in \mathbb{R}^{N} $}
\KwOut{Final Reconstruction: $ X^{*} $}
$ \mathbf{X} \leftarrow \mathbf{W} \leftarrow
\left(\begin{smallmatrix} 
X^{(0)} \\
\vdots \\
X^{(0)}
\end{smallmatrix}\right) $ \;

\While{not converged}
{
    $ \mathbf{X} \leftarrow \mathbf{\tilde{L}}(\mathbf{W};\mathbf{X}) $\;
    $ \mathbf{Z} \leftarrow \mathbf{G}( 2 \mathbf{X} - \mathbf{W}) $\;
    $ \mathbf{W} \leftarrow \mathbf{W} + 2 \rho(\mathbf{Z} - \mathbf{X}) $
}
$ X^{*} \leftarrow X_1 $

\caption{Fusing denoisers using MACE}
\end{algorithm}

\begin{figure}[ht]
\centering     
\includegraphics[width=0.50\textwidth]{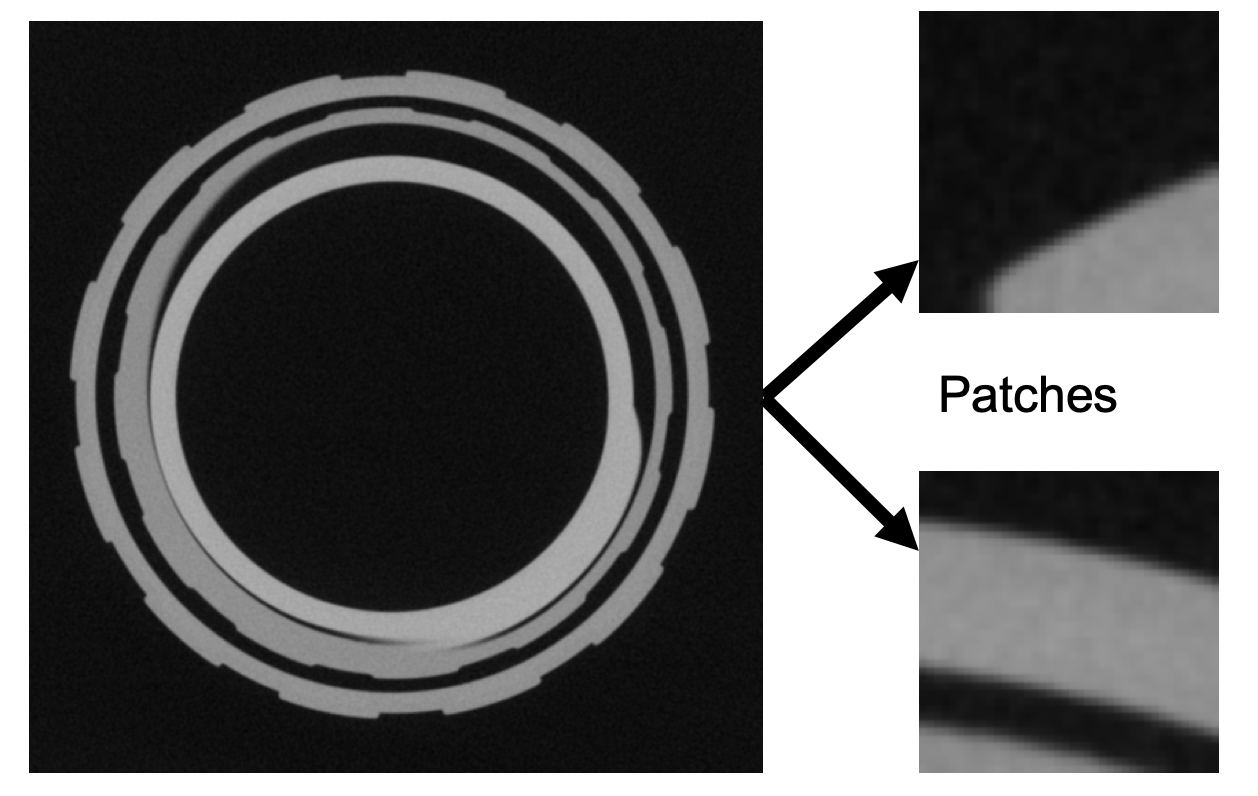}
\caption{Illustration of our training volume. We extract 3D patches from the volume to train our network.}
\label{fig:train}
\end{figure}

\section{CNN Space-Time Denoiser}
\label{sec:denoiser}
Since we iteratively apply multiple denoisers to multiple 4D volumes to compute the reconstruction, it is essential for each denoiser to be fast.
We use Deep Convolutional Neural Networks since they provide state-of-the-art quality in denoising while at the same time being fast due to optimized GPU libraries.

The architecture of the CNN denoiser is shown in Figure~\ref{fig:CNN}.
Each of the CNN denoisers used in the Multi-Slice Fusion is designed to perform 2.5D denoising of a 3D volume using a 2D moving window. 
By 2.5D we mean that as an input the neural network takes a thin 3D patch consisting of five consecutive 2D slices while the output is the denoised center slice.
The other slices are being denoised by shifting the 5-slice moving window.

The 2D slices are processed using 2D convolution while the adjacent slices are treated as separate input channels.
The neural network architecture is modeled after the residual network architecture presented in \cite{dncnn} but extended so that it works on 3D rather than 2D images.
In \cite{ziabariCNN,jiang2018denoising}, it has been shown that this type of 2.5D processing is a computationally efficient way of performing effective 3D denoising with CNNs. We note that it is similar to the method presented in \cite{dncnn} where the 5 adjacent slices are analogous to the 3 color channels used in denoising of color images. 


Our 3D CNN denoiser can theoretically be trained with space-time data to get different time and space regularization.
However, in the absence of space-time training data, we trained the network on 3D-space volumes.
For training, we used a CT reconstruction of a bottle and screw cap made from different plastics that has low noise.
A slice from the training volume is illustrated in Figure~\ref{fig:train}.
While our training and testing volumes are different, all of them are representative of a variety of NDE problems in which objects to be imaged are constructed from a relatively small number of distinct materials.

We extracted $63,000$ partially overlapping 3D patches of size $40 \times 40 \times 5$ from our training volume of size $534 \times 546 \times 100$.
As a data augmentation technique random rotation, flipping, and mirroring were applied during training.
Pseudo-random additive white Gaussian noise (AWGN) was applied to the patches and the network was trained to remove the noise. 
The AWGN noise used does not represent CT measurement noise but is a result of using the MACE framework.
More specifically, the AWGN noise that the denoisers removes in MACE is an 'artificial' noise induced from the quadratic norm used in equation~(\ref{eq:F_def}).

\section{4D Prior as Multi-Slice Fusion}

In this section we describe how we fuse the space-time denoisers described in section~\ref{sec:denoiser} using the approach in section~\ref{sec:MACE_fusion} to form a 4D reconstruction.

Although, the approach described in section~\ref{sec:MACE_fusion} can fuse arbitrary number of denoisers, we fuse three 3D CNN denoisers to form our 4D prior.
We refer to denoisers $H_1, H_2, H_3$ in equation~(\ref{eq:L_operator}) as $H_{xy,t}$, $H_{yz,t}$, and $H_{zx,t}$, respectively.
The denoiser $H_{xy,t}$ is a CNN space-time denoiser that does convolution in the xy plane and uses five input channels to input slices from neighboring time-points.
The denoisers $H_{yz,t}$ and $H_{zx,t}$ are analogous to $H_{xy,t}$ but are applied along the yz and zx plane, respectively. Each of the denoisers, despite being 3D denoisers, denoise the 4D volume by processing ``slice by slice''.

Each denoiser incorporates information along a spatial slice and uses a moving temporal window of 5 slices to denoise a single slice that is in the center.
However, our Multi-Slice Fusion fuses all the denoisers $H_{xy,t}$, $H_{yz,t}$, and $H_{zx,t}$ to incorporate information along all four dimensions.

\section{Experimental Results}

To demonstrate the improved reconstruction quality of our method, we present results on both real and simulated 4D X-ray CT data for NDE applications.
In both cases we compare Multi-Slice Fusion reconstruction with 4D Markov Random Field based MBIR reconstruction (MBIR+4D-MRF) and conventional 3D Filtered Back Projection reconstruction (FBP).
The MBIR+4D-MRF used was a variant of the QGGMRF model described in \cite{mohan2015timbir} with 26 spatial and 2 temporal neighbors.
We also compare with reconstructions using single CNNs to illustrate the effect of model fusion.
We refer to reconstructions using single CNN denoisers $H_{xy,t}$, $H_{yz,t}$, and $H_{zx,t}$ as MBIR+$H_{xy,t}$, MBIR+$H_{yz,t}$, and MBIR+$H_{zx,t}$, respectively.

In all our experiments the cone-beam inversion within Multi-Slice Fusion was executed on two CPU cluster nodes, each with 20 Kaby Lake CPU cores.
It was parallelized over time-points trivially and within time points using multiple threads similar to \cite{balke2018separable}.
We perform three equits per each Mann iteration for the cone-beam inversion.
The CNN denoisers ran in parallel with the cone-beam inversion in a separate cluster with a Nvidia Tesla P100 GPU.

\subsection{Real Dataset}

\begin{table}
\centering{} 
\small
\begin{tabular}{r|l}
\toprule
Scanner Model & \begin{tabular}{@{}l@{}}North Star Imaging \\ \textit{(Rogers, MN) X50}\end{tabular}   \\
Source-Detector Distance & 839 $\mathrm{mm}$  \\
Magnification & 5.57 \\
Number of Views per Rotation & 150 \\
Cropped Detector Array & $731 \times 91$, $(0.25 \ \mathrm{mm})^2$\\
Voxel Size & $(0.0456 \ \mathrm{mm})^3$ \\
Reconstruction Size $(x,y,z,t)$ & $731 \times 731 \times 91 \times 16 $ \\
\bottomrule
\hline
\end{tabular}
\\
\vspace{1mm}
\caption{Experimental setup for Real X-ray Data}
\label{table:setup}
\end{table}

In this section we present results on real data to evaluate our method.
The data is from a dynamic cone-beam X-ray scan of a glass vial, with elastomeric stopper and aluminum crimp-seal, using a North Star Imaging (Rogers, MN) X50 X-ray CT system.
The vial is undergoing dynamic compression during the scan, to capture the mechanical response of the components.
Of particular interest is the moment when the aluminum seal is no longer in contact with the underside of the glass neck finish.
This indicates the moment when the force applied exceeds that exerted by the rubber on the glass; this is known as the ``residual seal force'' \cite{vialCap}.
During the scan the vial was held in place by fixtures going in and out of the Field of View thus resulting in artifacts.
All our 4D results incorporate preprocessing to correct these artifacts.
The experimental setup is summarized in Table~\ref{table:setup}.

\begin{figure}[ht]
\centering     
\subfigure[FBP (3D)]{\includegraphics[width=0.15\textwidth]{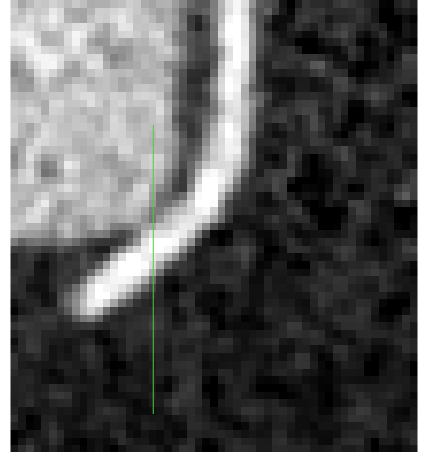}
\label{fig:results_crossSection:FBP}}
\subfigure[MIBIR+4D-MRF]{\includegraphics[width=0.15\textwidth]{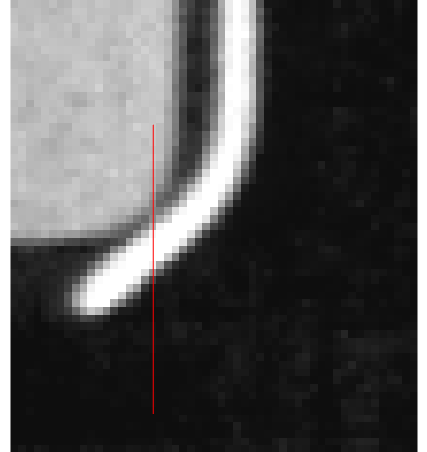}
\label{fig:results_crossSection:4D-MRF}}
\subfigure[Multi-Slice Fusion]{\includegraphics[width=0.15\textwidth]{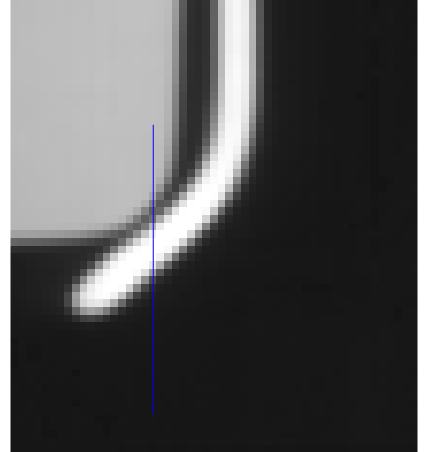}
\label{fig:results_crossSection:Our}}
\subfigure[Plot of cross-section]{\includegraphics[width=0.5\textwidth]{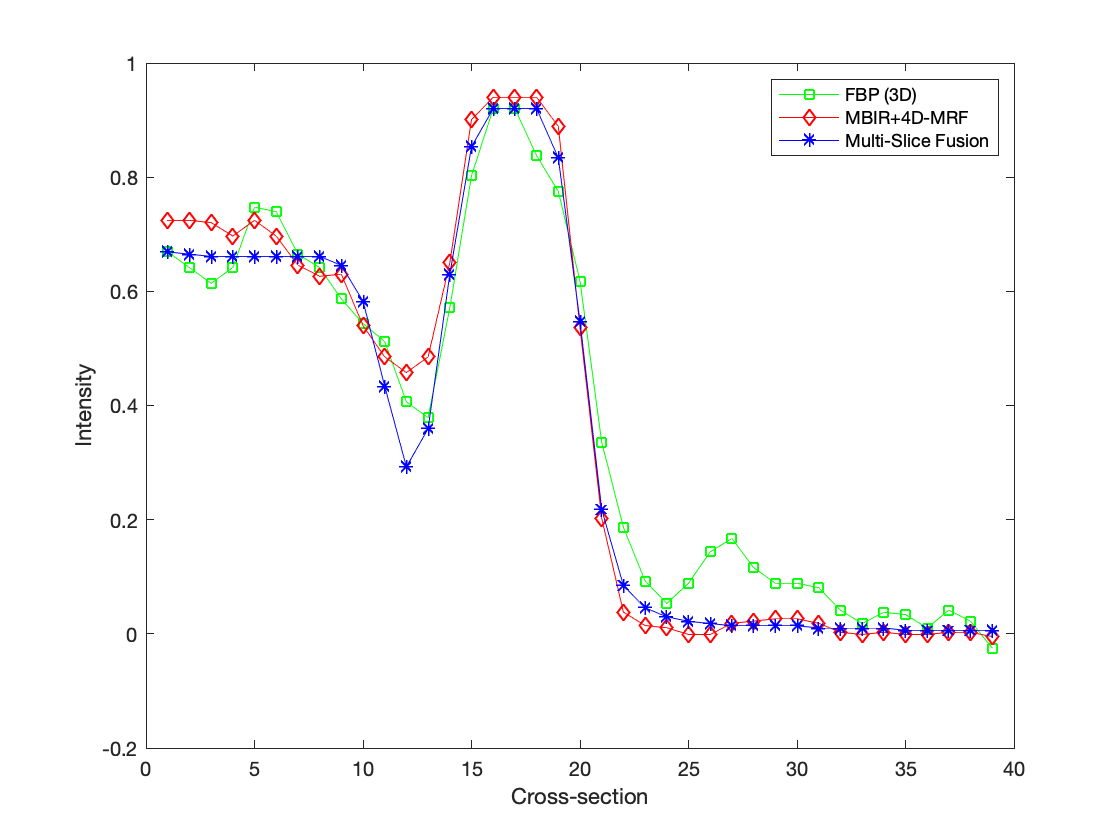}
\label{fig:results_crossSection:plot}}
\caption{ Plot of cross-section through the vial at a time when the aluminum and glass have physically separated.
Multi-Slice Fusion is able to resolve the junction between materials better while simultaneously producing a smoother reconstruction within materials compared to MBIR+4D-MRF and FBP.}
\label{fig:results_crossSection}
\end{figure} 

\begin{figure}[ht]
\centering     
\includegraphics[width=0.50\textwidth]{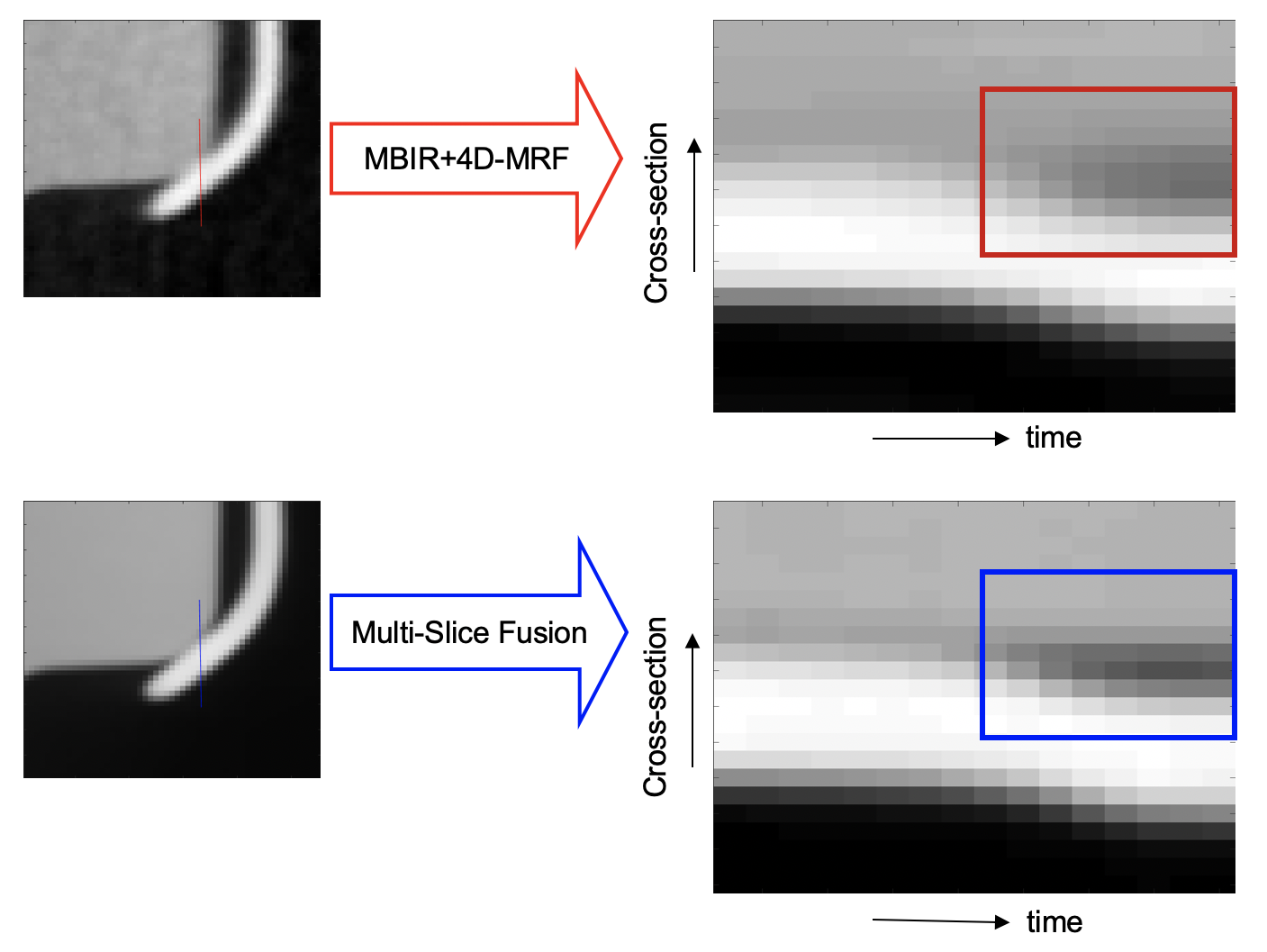}
\caption{Illustration of temporal resolution.
Multi-Slice Fusion results in improved space-time resolution of the separation of aluminum and glass.}
\label{fig:time_resolution}
\end{figure}

\begin{figure*}
\centering     

\subfigure[FBP (3D)]{\includegraphics[trim={0.9cm 0 0 0},clip,width=0.32\textwidth]
{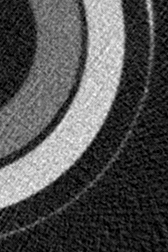}}
{\label{fig:results_xy:FPB}}
\subfigure[MBIR+4D-MRF]{\includegraphics[trim={0.9cm 0 0 0},clip,width=0.32\textwidth]
{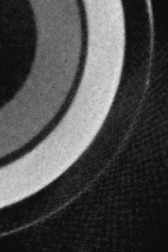}}
{\label{fig:results_xy:4D-MRF}}
\subfigure[Multi-Slice Fusion]{\includegraphics[trim={0.9cm 0 0 0},clip,width=0.32\textwidth]
{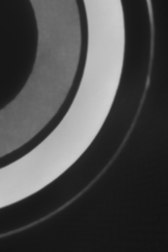}}
{\label{fig:results_xy:Our}}
\subfigure[MBIR+$H_{xy,t}$ (Missing Feature)]{\includegraphics[trim={0.9cm 0 0 0},clip,width=0.32\textwidth]
{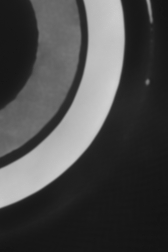}}
{\label{fig:results_xy:xy}}
\subfigure[MBIR+$H_{yz,t}$ (Horizontal Streaks)]{\includegraphics[trim={0.9cm 0 0 0},clip,width=0.32\textwidth]
{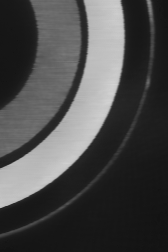}}
{\label{fig:results_xy:yz}}
\subfigure[MBIR+$H_{zx,t}$ (Vertical Streaks)]{\includegraphics[trim={0.9cm 0 0 0},clip,width=0.32\textwidth]
{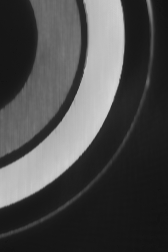}}
{\label{fig:results_xy:zx}}
\caption{ Comparison of different methods.
Each image is a slice through the reconstructed vial for one time point along the spatial xy plane.
Both FBP and MBIR+4D-MRF suffer from obvious windmill artifacts, higher noise and have blurred contours.
In contrast to that, the Multi-Slice Fusion reconstruction has smooth and uniform textures while preserving edge definition.
MBIR+$H_{yz,t}$ and MBIR+$H_{zx,t}$ suffer from horizontal and vertical streaks.
MBIR+$H_{xy,t}$ cannot reconstruct the outer ring since the slice displayed is at the edge of the aluminum seal and the xy plane does not contain sufficient information.
Multi-Slice Fusion can resolve the edges of the rings better than either of MBIR+$H_{xy,t}$, MBIR+$H_{yz,t}$, and MBIR+$H_{zx,t}$ since it has information from all the spatial coordinates.
}
\label{fig:results_xy}
\end{figure*}

Figure~\ref{fig:results_xy} compares Multi-Slice Fusion with several other methods.
Each image is a slice through the reconstructed vial for one time point along the spatial xy plane.
Both FBP and MBIR+4D-MRF suffer from obvious windmill artifacts, higher noise and have blurred contours.
In contrast to that, the Multi-Slice Fusion reconstruction has smooth and uniform textures while preserving edge definition.
Figure~\ref{fig:results_xy} also illustrates the effect of model fusion by comparing Multi-Slice Fusion with MBIR+$H_{xy,t}$, MBIR+$H_{yz,t}$, and MBIR+$H_{zx,t}$.
MBIR+$H_{yz,t}$ and MBIR+$H_{zx,t}$ suffer from horizontal and vertical streaks respectively since the denoisers were applied in those planes.
MBIR+$H_{xy,t}$ does not suffer from streaks in the figure since we are viewing a slice along the xy plane, but it suffers from other artifacts.
MBIR+$H_{xy,t}$ cannot reconstruct the outer ring since the slice displayed is at the edge of the aluminum seal and the xy plane does not contain sufficient information.
In contrast, Multi-Slice Fusion can resolve the edges of the rings better than either of MBIR+$H_{xy,t}$, MBIR+$H_{yz,t}$, and MBIR+$H_{zx,t}$ since it uses information from all the spatial coordinates.

Next, we plot a cross-section through the object for Multi-Slice Fusion, MBIR+4D-MRF and FBP in Figure~\ref{fig:results_crossSection}. 
For this, we choose a time-point where we know the aluminum and glass have separated spatially.
The Multi-Slice Fusion reconstruction has a steeper and more defined slope in the junction of aluminum and glass.
This supports that Multi-Slice Fusion is able to preserve fine details in spite of producing a smooth regularized image.

Finally in Figure~\ref{fig:time_resolution} we plot a cross-section through the object with time to show the improved space-time resolution of our method. We do this for both the 4D methods: MBIR+4D-MRF and Multi-Slice Fusion.
Multi-Slice Fusion results in improved space-time resolution of the separation of aluminum and glass.

\subsection{Simulated Data}

\begin{table}
\centering{} 
\small
\begin{tabular}{r|l}
\toprule
Source-Detector Distance & 839 $\mathrm{mm}$  \\
Magnification & 5.57 \\
Number of Views per Rotation & 75 \\
Cropped Detector Array & $240 \times 28$, $(0.25 \ \mathrm{mm})^2$\\
Voxel Size & $(0.0456 \ \mathrm{mm})^3$ \\
Reconstruction Size $(x,y,z,t)$ & $240 \times 240 \times 28 \times 8 $ \\
\bottomrule
\hline
\end{tabular}
\\
\vspace{1mm}
\caption{Experimental setup for Simulated X-ray Data}
\label{table:simsetup}
\end{table}

In this section we present results on simulated data to evaluate our method.
We take a low-noise CT reconstruction of a bottle and screw cap and denoise it further using BM4D~\cite{bm4d} to generate a clean 3D phantom.
We then move the 3D volume vertically for each time point to generate a 4D phantom.
We forward project the phantom to generate sinogram data and use that to reconstruct the object.
The simulated experimental setup is summarized in Table~\ref{table:simsetup}.

\begin{figure*}
\centering     
\subfigure[Phantom]{\includegraphics[trim={1.4cm 1.2cm 5.8cm 1.2cm},clip,width=0.13\textwidth]
{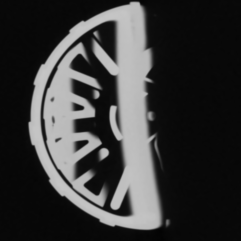}}
{\label{fig:simresults_xy:phantom}}
\subfigure[FBP (3D) ]{\includegraphics[trim={1.4cm 1.2cm 5.8cm 1.2cm},clip,width=0.13\textwidth]
{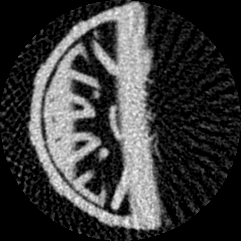}}
{\label{fig:simresults_xy:FPB}}
\subfigure[MBIR+4D-MRF ]{\includegraphics[trim={1.4cm 1.2cm 5.8cm 1.2cm},clip,width=0.13\textwidth]
{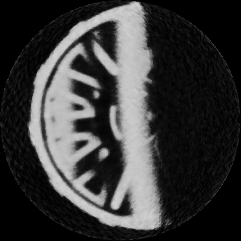}}
{\label{fig:simresults_xy:4D-MRF}}
\subfigure[Multi-Slice Fusion ]{\includegraphics[trim={1.4cm 1.2cm 5.8cm 1.2cm},clip,width=0.13\textwidth]
{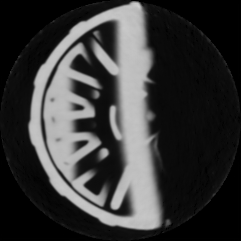}}
\hfill
{\label{fig:simresults_xy:Our}}
\subfigure[MBIR+$H_{xy,t}$]{\includegraphics[trim={1.4cm 1.2cm 5.8cm 1.2cm},clip,width=0.13\textwidth]
{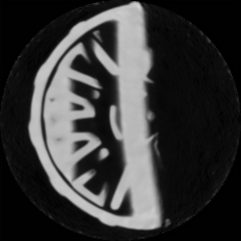}}
{\label{fig:simresults_xy:xy}}
\subfigure[MBIR+$H_{yz,t}$]{\includegraphics[trim={1.4cm 1.2cm 5.8cm 1.2cm},clip,width=0.13\textwidth]
{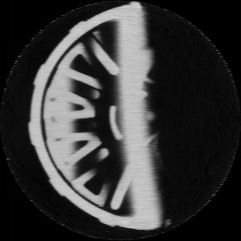}}
{\label{fig:simresults_xy:yz}}
\subfigure[MBIR+$H_{zx,t}$]{\includegraphics[trim={1.4cm 1.2cm 5.8cm 1.2cm},clip,width=0.13\textwidth]
{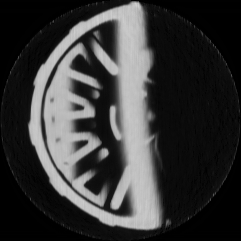}}
{\label{fig:simresults_xy:zx}}
\caption{ Comparison of different methods.
Each image is a slice through the reconstructed object for one time point along the spatial xy plane.
Both FBP and MBIR+4D-MRF suffer from high noise and jagged edges and fail to recover the small hole in the bottom of the image.
MBIR+$H_{yz,t}$ and MBIR+$H_{zx,t}$ suffer from horizontal and vertical streaks respectively since the denoisers were applied in those planes.
MBIR+$H_{xy,t}$ cannot reconstruct the small hole in the bottom of the image since the xy plane does not contain sufficient information.
}
\label{fig:simresults_xy}
\end{figure*}

Figure~\ref{fig:simresults_xy} compares Multi-Slice Fusion with several other methods.
Each image is a slice through the reconstructed object for one time point along the spatial xy plane.
Both FBP and MBIR+4D-MRF suffer from high noise and jagged edges and fail to recover the small hole in the bottom of the image.
MBIR+$H_{yz,t}$ and MBIR+$H_{zx,t}$ suffer from horizontal and vertical streaks respectively since the denoisers were applied in those planes.
MBIR+$H_{xy,t}$ does not suffer from streaks in the figure since we are viewing a slice along the xy plane, but it suffers from other artifacts.
MBIR+$H_{xy,t}$ cannot reconstruct the small hole in the bottom of the image since the xy plane does not contain sufficient information.

\begin{figure}[ht]
\centering     
\subfigure[Phantom]{\includegraphics[width=0.1\textwidth]{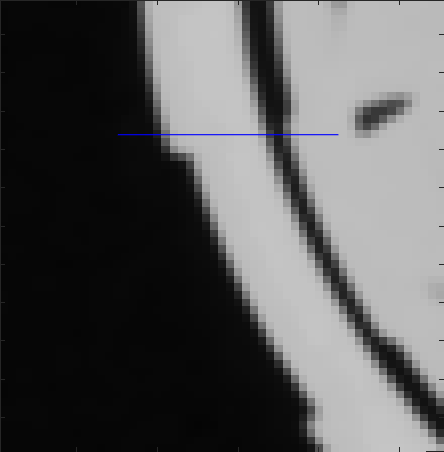}
\label{fig:results_crossSection:Phantom}}
\subfigure[FBP (3D)]{\includegraphics[width=0.1\textwidth]{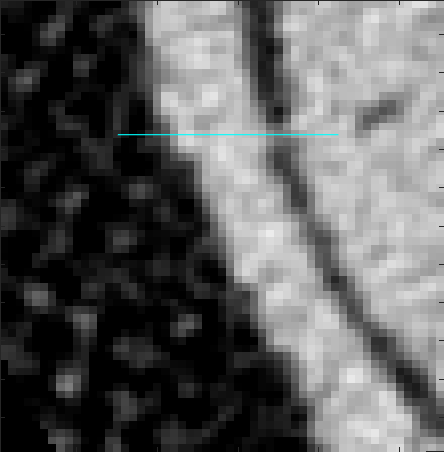}
\label{fig:simresults_crossSection:FBP}}
\subfigure[MIBIR+4D-MRF]{\includegraphics[width=0.1\textwidth]{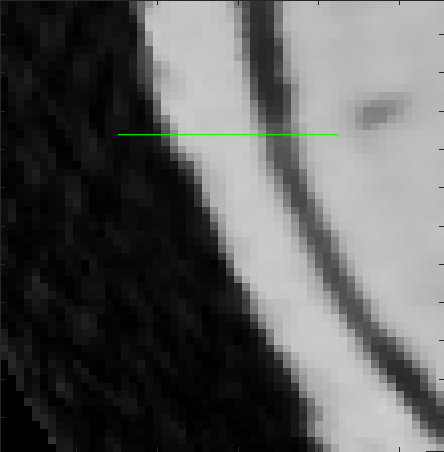}
\label{fig:simresults_crossSection:4D-MRF}}
\subfigure[Multi-Slice Fusion]{\includegraphics[width=0.1\textwidth]{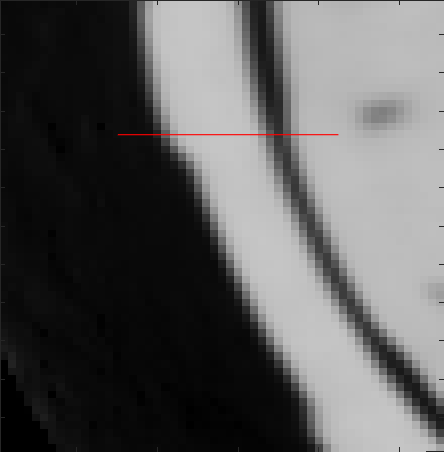}
\label{fig:simresults_crossSection:Our}}
\subfigure[Plot of cross-section]{\includegraphics[width=0.5\textwidth]{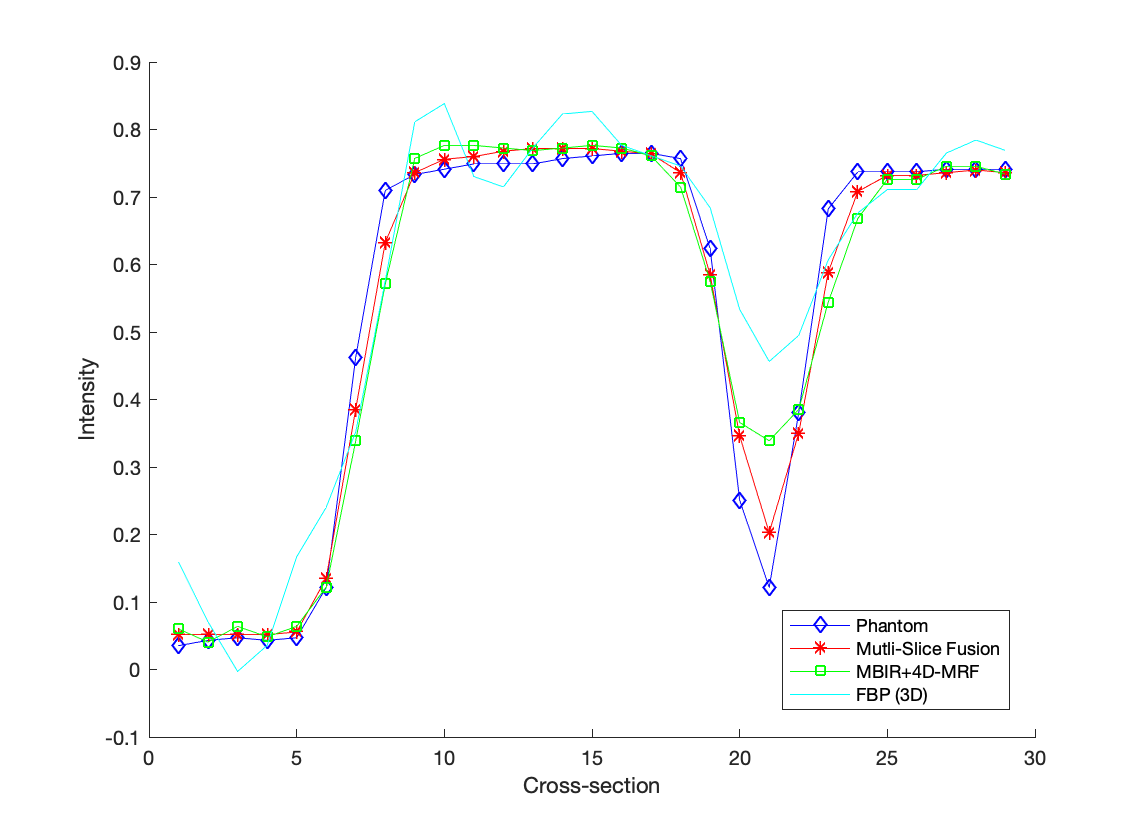}
\label{fig:simresults_crossSection:plot}}
\caption{ Plot of cross-section through the phantom and reconstructions from simulated data.
Multi-Slice Fusion results in the most accurate reconstruction of the gap between materials.}
\label{fig:simresults_crossSection}
\end{figure}

Next we plot a cross-section through the object for Multi-Slice Fusion, MBIR+4D-MRF, FBP, and the phantom Figure~\ref{fig:simresults_crossSection}.
Multi-Slice Fusion results in the most accurate reconstruction of the gap between materials.

Finally we report the Peak Signal to Noise Ratio (PSNR) and Structural Similarity Index (SSIM) with respect to the phantom for each method in Table~\ref{table:metrics} to objectively measure image quality.
We define the PSNR for a given reconstruction $X$ with a phantom $X^0$ as
\begin{equation}
    PSNR(X) = 20 \log_{10} \left( \dfrac{Range(X^0)}{RMSE(X,X^0)}
    \right) ,
\end{equation}
where range is computed from the $0.1$ and $99.9$ percentiles of the phantom.
As can be seen from Table~\ref{table:metrics}, Multi-Slice Fusion results in the highest PSNR and SSIM scores.

\begin{table}[ht]
\centering{} 
\small
\begin{tabular}{r|l|l}
\toprule
Method & PSNR(dB) & SSIM\\
\hline
FBP & 19.690 & 0.609 \\
MBIR+4D-MRF & 25.837 & 0.787 \\
Multi-Slice Fusion & \textbf{29.071} & \textbf{0.943} \\
MBIR+$H_{xy,t}$ & 29.026 & 0.922 \\
MBIR+$H_{yz,t}$ & 28.040 & 0.932 \\
MBIR+$H_{zx,t}$ & 28.312 & 0.926 \\
\bottomrule
\hline
\end{tabular}
\\
\vspace{1mm}
\caption{Quantitative Evaluation.
Multi-Slice Fusion has the highest PSNR and SSIM metric among all the methods.}
\label{table:metrics}
\end{table}

\section{Conclusion}

In this paper, we proposed a novel 4D X-Ray CT reconstruction algorithm, Multi-Slice Fusion, that combines multiple low dimensional denoisers to form a 4D prior.
Our method allows the formation of an advanced 4D prior using state-of-the-art CNN denoisers without needing to train on 4D data.
Furthermore, it allows for multiple levels of parallelism; thus enabling reconstruction of large volumes in a reasonable time. 
Although we focused on 4D X-Ray CT reconstruction for Non Destructive Evaluation (NDE) applications, our method can be used for any reconstruction problem involving multiple dimensions.

\ifpeerreview
\else
\section*{Acknowledgment}

The authors would like to acknowledge support from Eli Lilly and Company under research project funding agreement
17099289.
Charles A. Bouman and Gregery T. Buzzard were supported in part by NSF grant CCF-1763896.
We also thank M. Cory Victor and Dr. Coralie Richard from Eli Lilly for their assistance and guidance in setting up the residual seal force test experiment.
\fi

\bibliographystyle{IEEEtran}
\bibliography{references}

\begin{thebibliography}{10}
\providecommand{\url}[1]{#1}
\csname url@samestyle\endcsname
\providecommand{\newblock}{\relax}
\providecommand{\bibinfo}[2]{#2}
\providecommand{\BIBentrySTDinterwordspacing}{\spaceskip=0pt\relax}
\providecommand{\BIBentryALTinterwordstretchfactor}{4}
\providecommand{\BIBentryALTinterwordspacing}{\spaceskip=\fontdimen2\font plus
\BIBentryALTinterwordstretchfactor\fontdimen3\font minus
  \fontdimen4\font\relax}
\providecommand{\BIBforeignlanguage}[2]{{%
\expandafter\ifx\csname l@#1\endcsname\relax
\typeout{** WARNING: IEEEtran.bst: No hyphenation pattern has been}%
\typeout{** loaded for the language `#1'. Using the pattern for}%
\typeout{** the default language instead.}%
\else
\language=\csname l@#1\endcsname
\fi
#2}}
\providecommand{\BIBdecl}{\relax}
\BIBdecl

\bibitem{5D_huang2014mr}
C.~Huang, J.~L. Ackerman, Y.~Petibon, T.~J. Brady, G.~El~Fakhri, and J.~Ouyang,
  ``{MR}-based motion correction for {PET} imaging using wired active {MR}
  microcoils in simultaneous {PET-MR}: Phantom study,'' \emph{Medical physics},
  vol.~41, no.~4, 2014.

\bibitem{mohan2015timbir}
K.~A. Mohan, S.~Venkatakrishnan, J.~W. Gibbs, E.~B. Gulsoy, X.~Xiao,
  M.~De~Graef, P.~W. Voorhees, and C.~A. Bouman, ``{TIMBIR}: A method for
  time-space reconstruction from interlaced views.'' \emph{IEEE Trans.
  Computational Imaging}, vol.~1, no.~2, pp. 96--111, 2015.

\bibitem{ziabariCNN}
A.~Ziabari, D.~H. Ye, K.~D. Sauer, J.~Thibault, and C.~A. Bouman, ``2.5{D} deep
  learning for {CT} image reconstruction using a multi-{GPU} implementation,''
  in \emph{Signals, Systems, and Computers, 2018 52nd Asilomar Conference
  on}.\hskip 1em plus 0.5em minus 0.4em\relax IEEE, 2018.

\bibitem{gibbs2015three}
J.~Gibbs, K.~A. Mohan, E.~Gulsoy, A.~Shahani, X.~Xiao, C.~Bouman, M.~De~Graef,
  and P.~Voorhees, ``The three-dimensional morphology of growing dendrites,''
  \emph{Scientific reports}, vol.~5, p. 11824, 2015.

\bibitem{mohan20154d}
K.~A. Mohan, S.~Venkatakrishnan, J.~W. Gibbs, E.~B. Gulsoy, X.~Xiao,
  M.~De~Graef, P.~W. Voorhees, and C.~A. Bouman, ``{4D} model-based iterative
  reconstruction from interlaced views.'' in \emph{ICASSP}, 2015, pp. 783--787.

\bibitem{wang2016fast}
X.~Wang, K.~A. Mohan, S.~J. Kisner, C.~Bouman, and S.~Midkiff, ``Fast voxel
  line update for time-space image reconstruction,'' in \emph{2016 IEEE
  International Conference on Acoustics, Speech and Signal Processing
  (ICASSP)}.\hskip 1em plus 0.5em minus 0.4em\relax IEEE, 2016, pp. 1209--1213.

\bibitem{kisner2012model}
S.~J. Kisner, E.~Haneda, C.~A. Bouman, S.~Skatter, M.~Kourinny, and S.~Bedford,
  ``Model-based {CT} reconstruction from sparse views,'' in \emph{Second
  International Conference on Image Formation in X-Ray Computed Tomography},
  2012, pp. 444--447.

\bibitem{sauer1993local}
K.~Sauer and C.~Bouman, ``A local update strategy for iterative reconstruction
  from projections,'' \emph{IEEE Transactions on Signal Processing}, vol.~41,
  no.~2, pp. 534--548, 1993.

\bibitem{sreehari2016plug}
S.~Sreehari, S.~V. Venkatakrishnan, B.~Wohlberg, G.~T. Buzzard, L.~F. Drummy,
  J.~P. Simmons, and C.~A. Bouman, ``Plug-and-play priors for bright field
  electron tomography and sparse interpolation,'' \emph{IEEE Transactions on
  Computational Imaging}, vol.~2, no.~4, pp. 408--423, 2016.

\bibitem{venkatakrishnan2013plug}
S.~V. Venkatakrishnan, C.~A. Bouman, and B.~Wohlberg, ``Plug-and-play priors
  for model based reconstruction,'' in \emph{Global Conference on Signal and
  Information Processing (GlobalSIP), 2013 IEEE}.\hskip 1em plus 0.5em minus
  0.4em\relax IEEE, 2013, pp. 945--948.

\bibitem{sun2018online}
Y.~Sun, B.~Wohlberg, and U.~S. Kamilov, ``An online plug-and-play algorithm for
  regularized image reconstruction,'' \emph{arXiv preprint arXiv:1809.04693},
  2018.

\bibitem{kamilov2017plug}
U.~S. Kamilov, H.~Mansour, and B.~Wohlberg, ``A plug-and-play priors approach
  for solving nonlinear imaging inverse problems,'' \emph{IEEE Signal
  Processing Letters}, vol.~24, no.~12, pp. 1872--1876, 2017.

\bibitem{bm3d}
K.~Dabov, A.~Foi, V.~Katkovnik, and K.~Egiazarian, ``Image denoising by sparse
  3{D} transform-domain collaborative filtering,'' \emph{IEEE Transactions on
  image processing}, vol.~16, no.~8, pp. 2080--2095, 2007.

\bibitem{bm4d}
M.~Maggioni, G.~Boracchi, A.~Foi, and K.~Egiazarian, ``Video denoising using
  separable 4{D} nonlocal spatiotemporal transforms,'' in \emph{Image
  Processing: Algorithms and Systems IX}, vol. 7870.\hskip 1em plus 0.5em minus
  0.4em\relax International Society for Optics and Photonics, 2011, p. 787003.

\bibitem{buzzard2018plug}
G.~T. Buzzard, S.~H. Chan, S.~Sreehari, and C.~A. Bouman, ``Plug-and-play
  unplugged: Optimization-free reconstruction using consensus equilibrium,''
  \emph{SIAM Journal on Imaging Sciences}, vol.~11, no.~3, pp. 2001--2020,
  2018.

\bibitem{jiang2018denoising}
D.~Jiang, W.~Dou, L.~Vosters, X.~Xu, Y.~Sun, and T.~Tan, ``Denoising of 3{D}
  magnetic resonance images with multi-channel residual learning of
  convolutional neural network,'' \emph{Japanese journal of radiology},
  vol.~36, no.~9, pp. 566--574, 2018.

\bibitem{dncnn}
K.~Zhang, W.~Zuo, Y.~Chen, D.~Meng, and L.~Zhang, ``Beyond a gaussian denoiser:
  Residual learning of deep {CNN} for image denoising,'' \emph{IEEE
  Transactions on Image Processing}, vol.~26, no.~7, pp. 3142--3155, 2017.

\bibitem{sun2018plug}
Y.~Sun, B.~Wohlberg, and U.~S. Kamilov, ``Plug-in stochastic gradient method,''
  \emph{arXiv preprint arXiv:1811.03659}, 2018.

\bibitem{sun2018regularized}
Y.~Sun, S.~Xu, Y.~Li, L.~Tian, B.~Wohlberg, and U.~S. Kamilov, ``Regularized
  fourier ptychography using an online plug-and-play algorithm,'' \emph{arXiv
  preprint arXiv:1811.00120}, 2018.

\bibitem{balke2018separable}
T.~Balke, S.~Majee, G.~T. Buzzard, S.~Poveromo, P.~Howard, M.~A. Groeber,
  J.~McClure, and C.~A. Bouman, ``Separable models for cone-beam {MBIR}
  reconstruction,'' \emph{Electronic Imaging}, vol. 2018, no.~15, pp. 181--1,
  2018.

\bibitem{sridhar2018distributed}
V.~Sridhar, G.~T. Buzzard, and C.~A. Bouman, ``Distributed framework for fast
  iterative {CT} reconstruction from view-subsets,'' \emph{Electronic Imaging},
  vol. 2018, no.~15, pp. 102--1, 2018.

\bibitem{sridharDistributed_CT_TCI}
V.~Sridhar, X.~Wang, G.~T. Buzzard, and C.~A. Bouman, ``Distributed memory
  framework for fast iterative {CT} reconstruction from view-subsets using
  multi-agent consensus equilibrium,'' \emph{Manuscript in preparation for IEEE
  Transactions on Computational Imaging}, 2018.

\bibitem{vialCap}
R.~Mathaes, H.-C. Mahler, J.-P. Buettiker, H.~Roehl, P.~Lam, H.~Brown,
  J.~Luemkemann, M.~Adler, J.~Huwyler, A.~Streubel \emph{et~al.}, ``The
  pharmaceutical vial capping process: Container closure systems, capping
  equipment, regulatory framework, and seal quality tests,'' \emph{European
  Journal of Pharmaceutics and Biopharmaceutics}, vol.~99, pp. 54--64, 2016.

\end{thebibliography}

\end{document}